\documentclass[twocolumn]{aastex63}

\shorttitle{The Dark Planets of WASP-47}
\shortauthors{Stephen R. Kane et al.}

\begin{document}

\title{The Dark Planets of the WASP-47 Planetary System}

\author{Stephen R. Kane}
\affiliation{Department of Earth and Planetary Sciences, University of
  California, Riverside, CA 92521, USA}
\email{skane@ucr.edu}

\author{Tara Fetherolf}
\affiliation{Department of Physics and Astronomy, University of
  California, Riverside, CA 92521, USA}

\author{Michelle L. Hill}
\affiliation{Department of Earth and Planetary Sciences, University of
  California, Riverside, CA 92521, USA}


\begin{abstract}

Exoplanet discoveries have demonstrated a vast range of planetary
system architectures. The demographic of compact planetary systems are
especially interesting from the perspective of planetary formation and
the evolution of orbital dynamics. Another interesting demographic is
that of giant planets in eccentric orbits, since these planets have
likely had a dynamical history involving planet-planet scattering
events. The WASP-47 system is particularly fascinating since it
combines these two demographics, having both compact planetary orbits
and a giant planet on an eccentric orbit within the system Habitable
Zone. Here we provide an analysis of the WASP-47 system from the
perspective of atmospheric detection and characterization. We discuss
the system architecture and the potential for additional long-period
planets. We simulate expected phase variations as a function of planet
orbital phase for the system due to the combined effect of the
planets. We present an analysis of precision photometry of WASP-47 from
the {\it K2} mission, phased on each of the planets. The analysis
rules out the detection of phase signatures for the two inner-most
planets, enabling constraints upon their albedos and atmospheric
properties. Our study concludes that WASP-47b is an example of a
``dark'' planet with a tentative geometric albedo of 0.016 and a
1$\sigma$ upper limit of 0.17. The WASP-47e data are consistent with a
broad range of albedos, but also show early evidence of having a
relatively low albedo. The growing number of dark, short-period giant
planets provide the framework of an ideal sample for studying low
albedo dependence on atmospheric composition.

\end{abstract}



\section{Introduction}
\label{intro}

A substantial number of compact planetary systems
\citep{funk2010,hands2014} have been discovered, particularly using
the transit method which is biased towards the compact system
detection space \citep{kane2008b}. Compact systems allow the
determination of the planet masses via Transit Timing Variations
(TTVs) due to the measurable dynamical interaction of the planets over
short time scales \citep{agol2005,holman2005}. Beyond the compact
system regime, a plethora of orbital architectures have been unveiled
\citep{winn2015,hatzes2016d}, including planets on highly eccentric
orbits \citep{kane2012d}. The formation and evolution of these various
architectures remains the subject of ongoing research, including
migration scenarios \citep{ford2014}, atmospheric response to variable
flux \citep{kane2017d}, and impacts on potential habitability
\citep{wolf2017b}. Detection of the atmospheres in systems with
diverse architectures could yield promising clues towards the nature
of formation processes \citep{madhusudhan2017}.

A diverse representation of orbital architectures may be found in the
WASP-47 planetary system. The WASP-47 system was discovered by
\citet{hellier2012}, with the detection of a short-period jovian
planet. A further two transiting planets were discovered via {\it K2}
observations \citep{becker2015b}; an ultra-short-period super-Earth
and a longer period Neptune-size planet. Radial velocity (RV)
monitoring of the system revealed a non-transiting giant planet in an
eccentric orbit \citep{neveu2016}. The masses of the transiting
planets have been determined through a combination of RVs
\citep{dai2015,sinukoff2017a,vanderburg2017,weiss2017}, TTVs
\citep{dai2015,weiss2017}, and photodynamical constraints
\citep{almenara2016}. The system is fascinating for numerous reasons,
including the diverse architecture, the potential for atmospheric
detection of the planets via phase variations as a function of planet
orbital phase, and the presence of an eccentric giant planet interior
to the snow line. The combination of these aspects allows for a deeper
understanding of the system history and provides significant
motivation for follow-up observations.

Here, we detail an analysis of both the WASP-47 system architecture
and precision photometry of the system from the {\it K2} mission
\citep{howell2014}. In Section~\ref{system}, we provide details of the
architecture including a discussion of the giant planet in the system
Habitable Zone (HZ) and evidence towards further long-period
planets. Section~\ref{phase} contains the results of a phase variation
simulation for the WASP-47 system, showing the expected relative
amplitude of the known planets and predictions regarding their
detectability. In Section~\ref{k2} we describe reprocessed photometry
from the {\it K2} mission that has been optimized towards analysis of
out-of-transit variability, from which we rule out significant phase
signatures of the two inner-most planets, constraining their albedos.
We discuss the implications of the low albedos on the
atmospheric/surface properties of the planets in
Section~\ref{discussion}, along with the potential for additional
planets and further follow-up opportunities.  We provide concluding
remarks and additional follow-up suggestions in
Section~\ref{conclusions}.


\section{System Architecture and Habitable Zone}
\label{system}

\begin{figure}
  \begin{center}
    \includegraphics[angle=270,width=8.5cm]{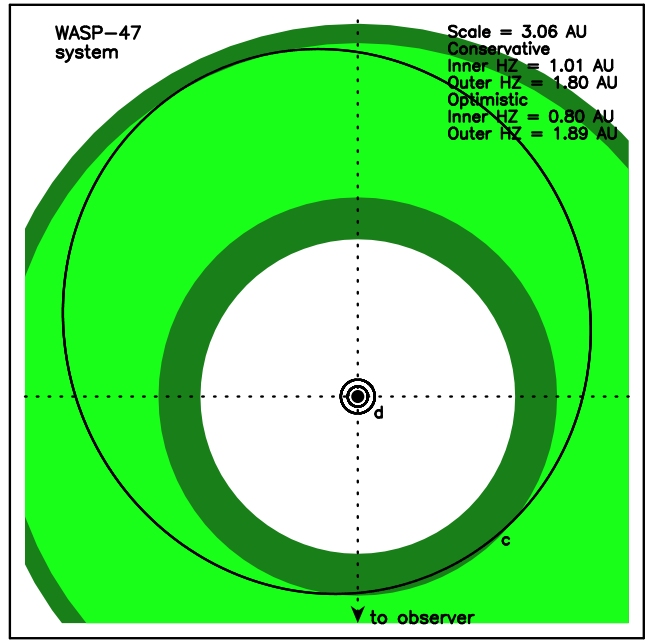}
  \end{center}
  \caption{A top-down view of the WASP-47 system, showing the host
    star (intersection of the dotted cross-hairs) and the orbits of
    the planets (solid lines). The extent of the HZ is shown in green,
    where light green is the conservative HZ and dark green is the
    optimistic extension to the HZ.}
  \label{systemfig}
\end{figure}

As described in Section~\ref{intro}, the current architecture of the
WASP-47 system has been revealed in stages that include both transit
and RV detection using both ground and space-based facilities. The
system includes three inner transiting planets and an outer
non-transiting planet in an eccentric ($e \sim 0.3$) orbit. The most
recent system parameters, provided by \citet{vanderburg2017}, are
shown in Table~\ref{paramtab} and for each planet include the orbital
period ($P$), semi-major axis ($a$), eccentricity ($e$), and
periastron argument ($\omega$), orbital inclination ($i$), radius
($R_p$), and mass ($M_p$). Note that we have assumed a Jupiter radius
for the non-transiting giant planet WASP-47c. WASP-47 is a star that
is similar to the Sun, with a mass of $M_\star =
1.040\pm0.031$~$M_\odot$, radius of $R_\star =
1.137\pm0.013$~$R_\odot$, and effective temperature of $T_\mathrm{eff}
= 5552\pm75$~K \citep{vanderburg2017}. A top-down representation of
the orbital architecture for the WASP-47 system is shown in
Figure~\ref{systemfig}.

\begin{deluxetable*}{cccccccccccccc}
  \tablecolumns{13}
  \tablewidth{0pc}
  \tablecaption{\label{paramtab} Planetary Parameters$^1$ and
    Derived Characteristics}
  \tablehead{
    \colhead{Planet} &
    \colhead{$P$} &
    \colhead{$a$} &
    \colhead{$e$} &
    \colhead{$\omega$} &
    \colhead{$i$} &
    \colhead{$R_p$} &
    \colhead{$M_p$} &
    \colhead{$T_\mathrm{eq}$} &
    \colhead{$f_\mathrm{IR}$} &
    \multicolumn{3}{c}{Optical Flux Ratio$^2$ (ppm)} \\
    \colhead{} &
    \colhead{(days)} &
    \colhead{(AU)} &
    \colhead{} &
    \colhead{(deg)} &
    \colhead{(deg)} &
    \colhead{($R_\oplus$)} &
    \colhead{($M_\oplus$)} &
    \colhead{(K)} &
    \colhead{(\%)} &
    \colhead{Rocky$^4$} &
    \colhead{Molten$^4$} &
    \colhead{Atmosphere$^3$}
  }
  \startdata
  e & 0.789592   & 0.0169 & 0.0   & --    & 85.98 & 1.810        & 6.83  & 2608 & 12.86 & 2.056 & 12.338 & 6.169 \\
  b & 4.1591289  & 0.0513 & 0.0   & --    & 88.98 & 12.63        & 363.1 & 1499 &  0.58 & --   & --     & 32.963 \\
  d & 9.03077    & 0.0860 & 0.0   & --    & 89.32 & 3.576        & 13.1  & 1158 &  0.05 & --   & --     & 0.940 \\
  c & 588.5      & 1.393  & 0.296 & 112.4 & --    & --$^4$ & 398.2 &  288 &  0.00 & --   & --     & 0.004 \\
  \enddata
  \tablenotetext{1}{From Table 3 of \citet{vanderburg2017}.}
  \tablenotetext{2}{Optical flux ratio between planet and star (see
    Equation~\ref{fluxratio}).}
  \tablenotetext{3}{"Rocky", "molten", and "atmosphere" models are
    equivalent to $A_g$ values of 0.1, 0.6, and 0.3 respectively (see
    Section~\ref{phase}).}
  \tablenotetext{4}{Assuming 1.0~$R_J$ = 11.2~$R_\oplus$ for planet c.}
\end{deluxetable*}

Using the above stellar parameters, we calculate the extent of the HZ
for the system. We adopt the HZ as described by
\citet{kopparapu2013a,kopparapu2014}, including the conservative and
optimistic boundaries \citep{kane2013d,kane2016c}. Briefly, the
demarcation of the conservative HZ is defined by the runaway
greenhouse limit at the inner edge and maximum greenhouse at the outer
edge, whereas the boundaries of the optimistic HZ are estimated based
on empirical evidence regarding the prevalence of liquid water on the
surfaces of Venus and Mars, respectively. Furthermore, the
uncertainties in the HZ boundaries depend on the robustness of the
stellar property values \citep{kane2014a}. As pointed out by
\citet{vanderburg2017}, the WASP-47 stellar properties are quite
accurately determined, aided substantially by the vast number of
measurements and similarity to the Sun. The representation of the
orbital architecture shown in Figure~\ref{systemfig} includes a green
region that indicates the extent of the HZ in the system
\citep{kane2012a}. The light green represents the conservative HZ and
the dark green represents the optimistic extension to the HZ. The
conservative and optimistic HZ span the ranges 1.01--1.80~AU and
0.80--1.89~AU respectively.

Figure~\ref{systemfig} indicates that, although in an eccentric orbit,
the non-transiting giant planet WASP-47c remains entirely within the
optimistic HZ during its orbit. Giant planets in the HZ are
intrinsically interesting from the perspective of potentially
habitable exomoons (see Section~\ref{discussion}). However, they also
pose intriguing dynamical questions regarding whether there could be
farther out giant planets that contributed to migration halting
mechanisms \citep{masset2001a,pierens2014b} and continuing dynamical
interactions \citep{kane2014b}. The current RV time series provided by
\citet{vanderburg2017} do not have a sufficient baseline to exclude
the presence of further long-period giant planets, the detection of
which would support the notion of dynamical interactions with a
farther companion, consistent with the eccentric giant planet detected
in the HZ.


\section{Planetary Phase Amplitudes}
\label{phase}

The photometric variations of a star--planet system due to the orbital
phase of the planet are dominated by three major effects: reflected
light and thermal emission, Doppler beaming, and ellipsoidal
variations \citep{faigler2011}. The precise nature of how these
various contributions to the photometric variations are weighted may
be used to discriminate between planetary and stellar companions
\citep{drake2003,kane2012b} and study multi-planet systems
\citep{kane2013b,gelino2014}. The Doppler beaming and ellipsoidal
variations are primarily functions of the companion mass
\citep{loeb2003,zucker2007b}, whereas the reflected light component
depends upon the planet radius and atmospheric properties
\citep{sudarsky2005} with further dependencies upon orbital parameters
\citep{kane2010b,kane2011a}. The precision of photometry from the {\it
  Kepler} spacecraft has demonstrated on numerous occasions that it is
sufficient to detect exoplanet phase variations
\citep{welsh2010,esteves2013,angerhausen2015b,esteves2015,shporer2015}.

\begin{figure*}
  \begin{center}
    \includegraphics[angle=270,width=16.0cm]{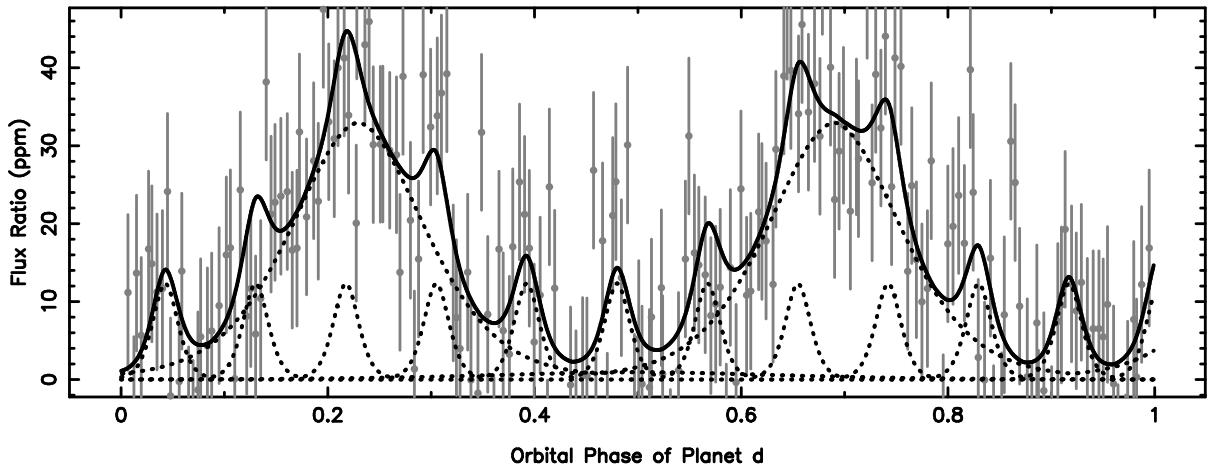}
  \end{center}
  \caption{Simulated phase variations for the WASP-47 system and for
    one complete orbit of planet d (9.03 days). The dotted lines
    represent the phase variations for the individual planets and the
    solid line is the combined phase effect. The gray data points are
    simulated photometric measurements to provide an estimate of
    expected signal with a gaussian scatter of 10~ppm.}
  \label{phasefig}
\end{figure*}

For the reflected light component observed at wavelength $\lambda$, we
adopt the formalism of \citet{kane2010b}. The star--planet separation
incorporating Keplerian orbital parameters is given by
\begin{equation}
  r = \frac{a (1 - e^2)}{1 + e \cos f}
  \label{separation}
\end{equation}
where $f$ is the true anomaly. The phase angle ($\alpha$) of the
planet is defined to be zero when the planet is located at superior
conjunction, and is given by
\begin{equation}
  \cos \alpha = - \sin (\omega + f)
  \label{phaseangle}
\end{equation}
To describe the scattering properties of atmospheres, we adopt a phase
function ($g(\alpha)$) that was empirically derived based upon {\it
  Pioneer} observations of Venus and Jupiter
\citep{hilton1992}. Combined with the geometric albedo of the
planetary surface ($A_g(\lambda)$), these parameters together
calculate the flux ratio ($\epsilon$) between the planet and star
\begin{equation}
  \epsilon(\alpha,\lambda) \equiv
  \frac{f_p(\alpha,\lambda)}{f_\star(\lambda)}
  = A_g(\lambda) g(\alpha,\lambda) \frac{R_p^2}{r^2}
  \label{fluxratio}
\end{equation}
where $f_p(\alpha,\lambda)$ and $f_\star(\lambda)$ are the fluxes
received by the observer from the planet and star respectively.

Observations and models show that planetary geometric albedos can span
a large range of values, depending on surface/atmosphere composition
and incident flux \citep{sudarsky2005,kane2010b,esteves2013}. For the
purposes of demonstration, we calculate planet--star flux ratios
(Equation~\ref{fluxratio}) for the WASP-47 planets using $A_g = 0.3$
to represent a typical atmospheric albedo based on the Solar System
planets. Since the inner planet (e) is likely terrestrial and exists
under extreme incident flux conditions, we also adopt $A_g = 0.1$ and
$A_g = 0.6$ for this planet to represent rocky (no atmosphere) and
molten surfaces respectively, based upon similar surfaces observed in
the Solar System and lava-ocean models
\citep{kane2011f,rouan2011}. The amplitude of the phase variations for
these assumptions are shown alongside the orbital parameters in
Table~\ref{paramtab}. The amplitudes demonstrate that, for the
reflected light component, planet b is expected to dominate the phase
variations and planets d and c are unlikely to contribute any
significant phase signature.

Shown in Figure~\ref{phasefig} is a model of the predicted phase
variations of the planets in the WASP-47 system for one complete
orbital period of planet d (9.03 days). The dotted lines indicate the
contributions from the individual planets and the solid line
represents the combined effects of all planets. In this model, all
planets start at inferior conjunction where the phase amplitude is
zero. We also adopt the atmosphere model for all planets except for
planet e, where a molten model has been adopted. The light gray data
points are simulated photometric measurements that have been convolved
with a gaussian filter of width 10~ppm to emulate the expected
dispersion of the data. As expected, the signature of planet b
dominates the variation in the photometry with planet e producing a
small but measurable effect.

It is worth considering the infrared (IR) component of the phase
variations due to the thermal emission of the planets. This IR
component is particularly important for modeling overall phase
variations when the planet is tidal locked, whereby the view of the
hot dayside varies with time \citep{kane2011g}. The IR contribution to
the total observed flux depends on the bandpass of the detector. For
the {\it Kepler} spacecraft, the detectors include sensitivity into IR
wavelengths, roughly spanning 420--900~nm \citep{borucki2010a}. To
estimate the IR flux contribution, we calculated the blackbody
equilibrium temperature ($T_\mathrm{eq}$) for each planet and the
associated flux. We then separately integrated the entire blackbody
curve and the region that falls within the {\it Kepler} bandpass,
allowing the calculation of the percentage thermal emission flux in
the bandpass ($f_\mathrm{IR}$). For example, the value of
$T_\mathrm{eq}$ for planet e is especially high (2608~K), resulting in
a substantial fraction of the thermal emission (12.86\%) falling
within the {\it Kepler} bandpass. Thus, for planet e, the IR component
may be a non-negligible amount of the photometric variations caused by
the planet. The $T_\mathrm{eq}$ and $f_\mathrm{IR}$ values for all
four WASP-47 planets are included in Table~\ref{paramtab}.


\section{Phase Curve Analysis of the {\it K2} Photometry}
\label{k2}

In this section we discuss our phase curve analysis and geometric
albedo measurements for the planets in the WASP-47 system. In
Section~\ref{k2data}, we explain how we manipulate the {\it K2}
photometry. Our phase curve model and the results of the phase curve
model fit to the {\it K2} photometry are given in
Section~\ref{phasemodel}. We derive upper limits on the geometric
albedo using an injection test described in Section~\ref{injtest} and
show the results of the injection test in Section~\ref{phaseresults}.
Finally, in Section~\ref{multiphase} we discuss how an individual
planet's significant phase curve may influence the phase curves of
other planets in the WASP-47 system.


\subsection{Data Treatment}
\label{k2data}

\begin{figure}
  \begin{center}
    \includegraphics[width=8.5cm]{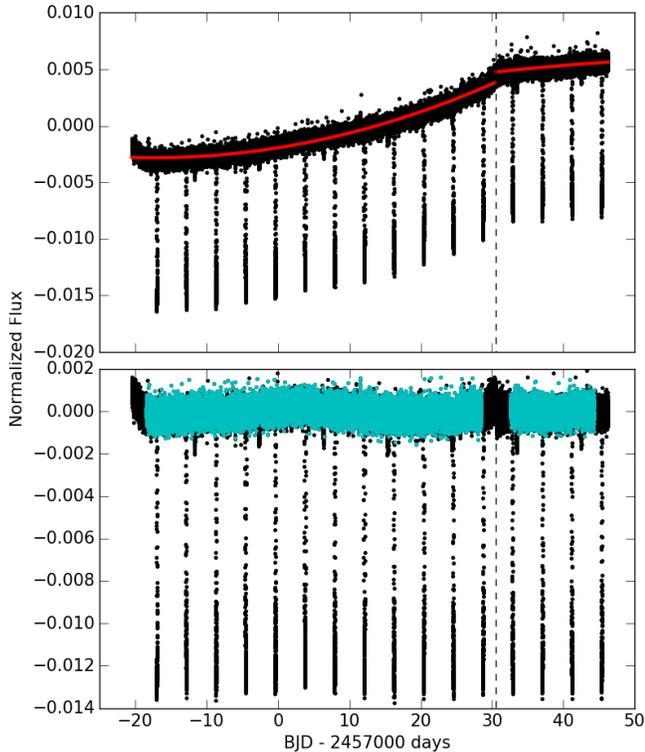}
  \end{center}
  \caption{Top: The {\it K2} photometry for WASP-47 after being
    corrected for systematic effects
    \citep{vanderburg2014,becker2015b}. The black vertical dashed line
    shows where we split the light curve into two segments, and the
    red curves show the best-fit 2nd-degree polynomial to the two
    non-transiting segments of the light curve. Bottom: The resultant
    light curve that has been normalized by the 2nd-degree polynomial
    in two segments. We use the light blue points in our phase
    variation analysis, which do not include transits or expected
    eclipses from any of the planets in WASP-47.}
  \label{lcfig}
\end{figure}

To analyze the phase curves of the planets in the WASP-47 system, we
use the short-cadence (58.3\,s) Pre-search Data Conditioned (PDC)
light curve of WASP-47 produced by the {\it Kepler/K2} pipeline. The
light curve is corrected for systematic effects, such as those caused
by unstable pointing, following the procedures described by
\citet{vanderburg2014} and \citet{becker2015b}, with the exception of
removing short frequency variations that are caused by the intrinsic
phase variations of the WASP-47 planets. However, some long-term
variability remains in the light curve, as can be seen in the top
panel of Figure~\ref{lcfig}. Therefore, we separate the light curve
into two segments, normalize each segment by the best-fit 2nd degree
polynomial, and remove two days of photometry at the edges of each
segment in order to avoid residual systematic effects caused by the
imperfect normalization that could bias our phase curve analysis. The
final light curve that we use in the phase variation analysis is shown
by the light blue points in the bottom panel of Figure~\ref{lcfig}.

\subsection{Phase Curve Fitting}
\label{phasemodel}

We use the times of conjunction, orbital periods, and transit duration
times reported in \citet{vanderburg2017} to phase fold the light curve
and trim the transits and expected eclipses for the b, d, and e
planets. WASP-47c is not included in our phase curve analysis given
that the available photometry only covers 11\% of its orbital
phase. Furthermore, the expected phase curve of WASP-47c is negligible
compared to the expected b, d, and e planet phase curves (see
Section~\ref{phase}). For the phase curve analysis, we adopt the
BEaming, Ellipsoidal, and Reflection (BEER) model developed by
\citet{faigler2011}. This model can be described as a simple, double
harmonic sinusoidal model, such that the flux ($F$) as a function of
phase ($\phi$) is given by
\begin{eqnarray}
  F(\phi) &=& A_\mathrm{norm} + A_\mathrm{refl} \cos{(2\pi\phi)}
  \nonumber \\ && + A_\mathrm{beam} \sin{(2\pi\phi)} +
  A_\mathrm{ellip} \cos{(4\pi\phi)} \text{,}
  \label{beereqn}
\end{eqnarray}
where $A_\mathrm{norm}$ is a normalization offset and
$A_\mathrm{beam}$, $A_\mathrm{ellip}$, and $A_\mathrm{refl}$ are the
semi-major amplitudes of the beaming, ellipsoidal, and reflection
effects, respectively\footnote{We also consider including additional
  sinusoidal terms to the model. However, the additional terms have
  negligible semi-major amplitudes, such that their inclusion does not
  significantly alter our results.}. The semi-amplitudes of the
beaming and ellipsoidal variations could be estimated and held fixed
in the model based on the known stellar and planetary parameters of
WASP-47 and its companions. However, we choose to leave these effects
as free parameters in the fitting due to the ellipsoidal distortion of
stars not being well understood \citep[e.g.,][]{shporer2017a} and the
beaming effect being degenerate with a phase-shifted reflection
component.  The signs of the amplitudes in Equation~\ref{beereqn} are
also allowed to be free in our fitting, but for the best-fit model to
be physically represented by the BEER model, the reflection and
ellipsoidal best-fit semi-amplitudes should be negative (equivalent to
a 180 degree phase shift).  The geometric albedo ($A_g$) in the {\it
  Kepler} bandpass can be measured for each planet using the strength
of the reflection component by
\begin{equation}
  A_g = \frac{A_{\text{refl}}}{\sin{i}} \left(\frac{R_p}{a}\right)^{-2}
  \text{.}
\label{albedoeqn}
\end{equation}
Since the b, d, and e planets all transit WASP-47, we consider the
effect of the inclination to be negligible and set $\sin{i}$ to
unity\footnote{Incorporating the inclinations from
  Table~\ref{paramtab} into our albedo measurements increases the
  albedo of WASP-47e by $10^{-3}$ and the albedos of the b and d
  planets by $10^{-4}$, which is less than the precision of the albedo
  measurements and thus does not alter our results.}.  A least-squares
minimization technique is used to find the best-fit BEER
model. Outlier removal is iterative, in that any data that are
$>$4.5$\sigma$ different from the model are removed, then the best-fit
BEER model is reevaluated. The remaining data are divided into 100
bins of the planet phase, then fit with the BEER model to ensure
consistency between the best-fit models.

\begin{figure}
  \begin{center}
    \includegraphics[clip,width=8.5cm]{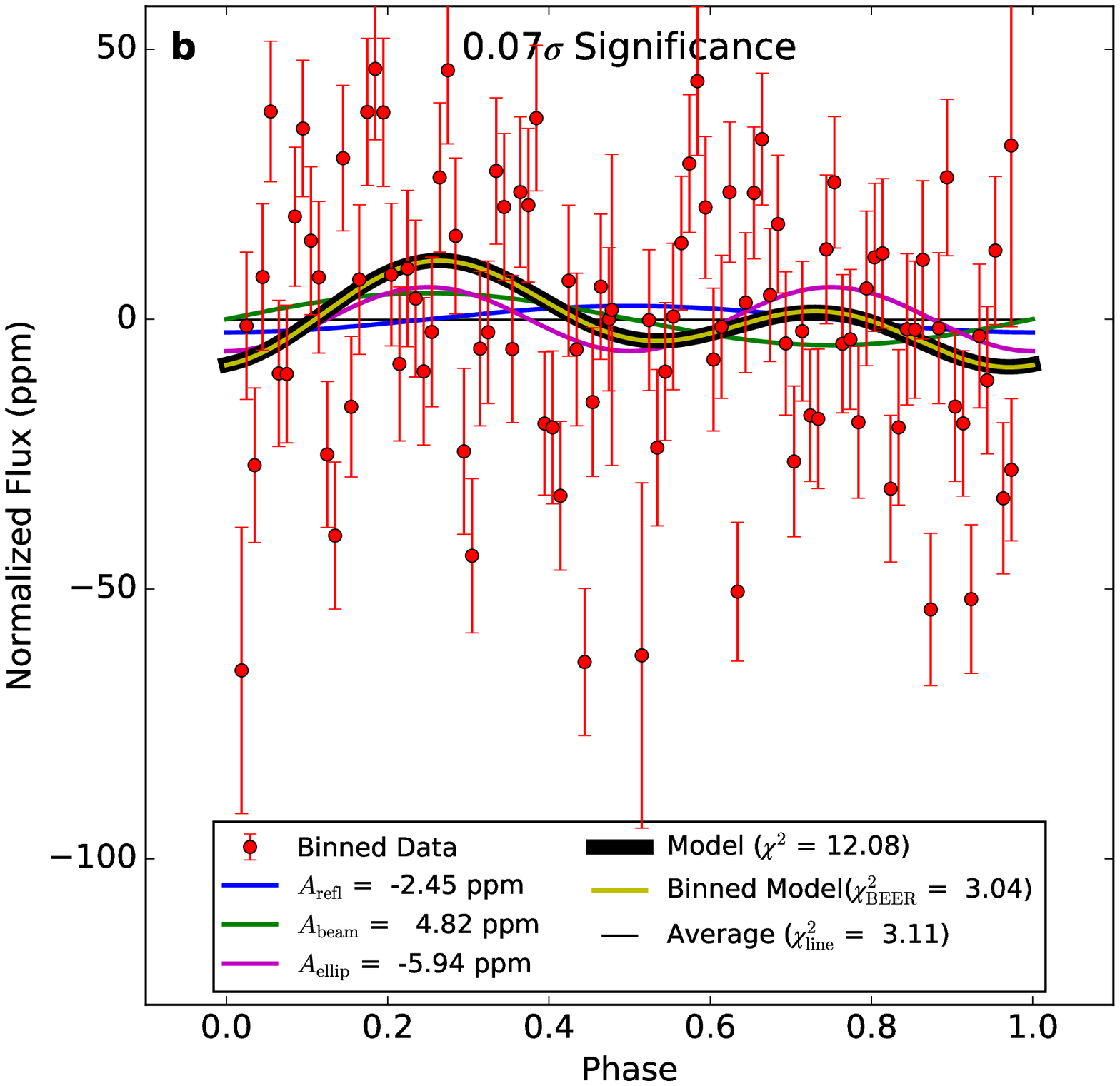} \\
    \includegraphics[clip,width=8.5cm]{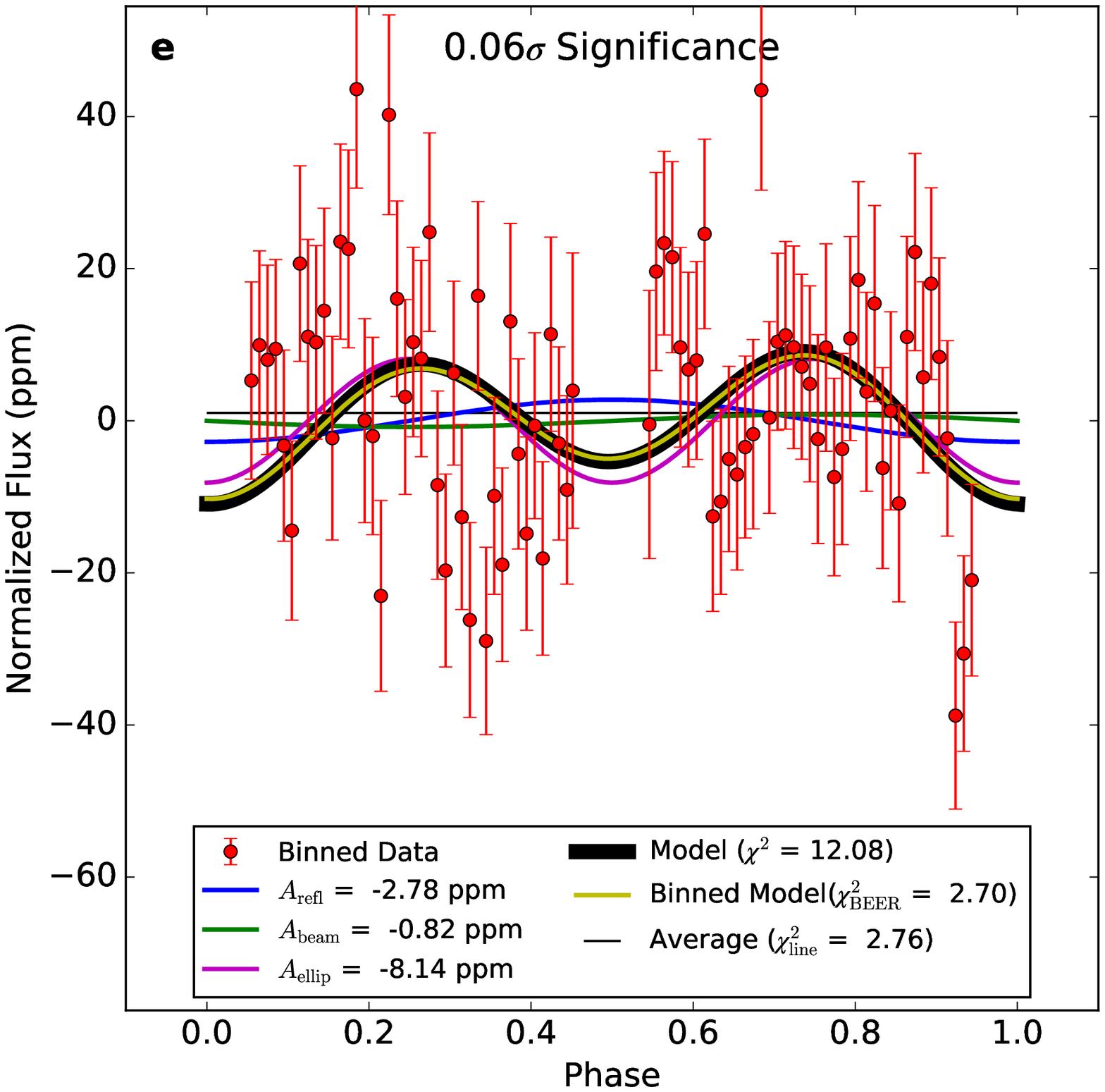}
  \end{center}
  \caption{Best-fit BEER model phase curves for the b (top row) and e
    (bottom row) planets in WASP-47, from which the phase curves are
    determined to be undetected (see Section~\ref{phasemodel}). The
    red points show the binned {\it K2} photometry. The thick black
    and thin yellow curves show the BEER models that are best-fit to
    the unbinned (not shown) and binned (red points) photometry,
    respectively. The individual components of the BEER model are
    shown by the blue (reflection), green (beaming), and purple
    (ellipsoidal) curves. The thin horizontal black line indicates the
    average flux of the data and is used to compare the phase curve to
    a flat line.  The significance of the best-fit model compared to a
    flat line is listed at the top of each panel.}
  \label{original}
\end{figure}

The best-fit BEER model is based on the shape of the light curve alone
and is not informed by the stellar or planetary parameters. The
stellar and planetary parameters are then typically used alongside the
measured semi-amplitudes of the BEER model to extract the planet's
albedo. We show the best-fit BEER model phase curves for the WASP-47 b
and e planets in Figure~\ref{original}. The best-fit phase curve of
WASP-47d is not shown since we find no evidence for BEER-related
effects based on the significance and signs of the best-fit
semi-amplitudes. We find that the measured phase curves of the b and e
planets do not significantly deviate from a flat line, therefore we
conclude that their phase curves are likely consistent with having a
very low albedo (Equation~\ref{albedoeqn}). In Section~\ref{injtest},
we aim to constrain the upper limit of the geometric albedos of the b
and e planets by injecting the expected phase curve model into the raw
{\it K2} photometry for different assumed albedos.

\subsection{Injection Test}
\label{injtest}

\begin{deluxetable*}{cccccccccccc}
  \tablecolumns{11}
  \tablewidth{0pc}
  \tablecaption{\label{injtab} Injected and Recovered Phase Variation
    Amplitudes$^\dagger$}
  \tablehead{
    \colhead{Planet} &
    \multicolumn{4}{c}{Injected Phase Curve} &
    \colhead{} &
    \multicolumn{5}{c}{Recovered Phase Curve} & \\
    \colhead{} &
    \colhead{$A_{\text{beam}}$} &
    \colhead{$A_{\text{ellip}}$} &
    \colhead{$A_{\text{refl}}$} &
    \colhead{$A_g$} &
    \colhead{} &
    \colhead{$A_{\text{refl}}$} &
    \colhead{$A_g$} &
    \colhead{$\chi^2_{\nu\text{,line}}$} &
    \colhead{$\chi^2_{\nu\text{,BEER}}$} &
    \colhead{$\chi^2_{\nu\text{,diff}}$} & \\
    \colhead{} & 
    \colhead{(ppm)} &
    \colhead{(ppm)} &
    \colhead{(ppm)} &
    \colhead{} & 
    \colhead{} &
    \colhead{(ppm)} &
    \colhead{} & 
    \colhead{} & 
    \colhead{} & 
    \colhead{} 
  }
  \startdata
  b & 1.68 & -1.50 & -14.76 & 0.17 & ~ & -19.12 & 0.18 & 4.09 & 3.08 & 1.01 \\
  ~ & ~ & ~ & -37.96 & 0.36 & ~ & -35.23 & 0.33 & 6.35 & 3.25 & 3.10 \\
  e & 0.06 & -0.78 & -13.50 & 0.68 & ~ & -12.61 & 0.64 & 3.01 & 2.51 & 0.5 \\
  ~ & ~ & ~ & -19.85 & 1.00 & ~ & -17.29 & 0.87 & 3.31 & 2.48 & 0.83 
  \enddata
  \tablenotetext{\dagger}{Results from the models corresponding to the
    phase curves shown in Figure~\ref{beerfig}.}
\end{deluxetable*}

Considering the nature of the systematic effects in the {\it K2}
photometry (e.g., reaction wheel jitter and solar pressure induced
drift), it is possible that the phase variations are removed when the
light curve is corrected. Therefore, we investigate whether a
BEER-like phase curve could be recovered when the predicted phase
variations are injected into the raw {\it K2} light curve for the
WASP-47 b, d, and e planets. We assume the stellar and planetary
parameters reported by \citet{vanderburg2017} and use the expressions
from \citet{faigler2011} to estimate the semi-amplitudes of the
beaming ($A_\mathrm{beam}$) and ellipsoidal ($A_\mathrm{ellip}$)
effects, although these signals are $<$2~ppm for all planets in the
WASP-47 system (see Table~\ref{injtab}) and is the same for all
injection tests.  The only difference between each injected phase
curve test was the strength of each planet's reflectivity. The
semi-amplitude of the reflection effect ($A_\mathrm{refl}$) for each
planet is determined by Equation~\ref{albedoeqn} in steps of 0.01
albedo. For each injection test, we add the combined phase curve
(Equation~\ref{beereqn}) of all three planets to the raw {\it K2}
photometry. For simplicity, we assume that all of the planets have the
same albedo for each injection test. We then repeat the {\it K2}
corrections described by \citet{vanderburg2014}, our 2nd degree
polynomial normalization on the two light curve segments, and our
phase curve analysis.

\begin{figure}
  \begin{center}
    \includegraphics[width=8.5cm]{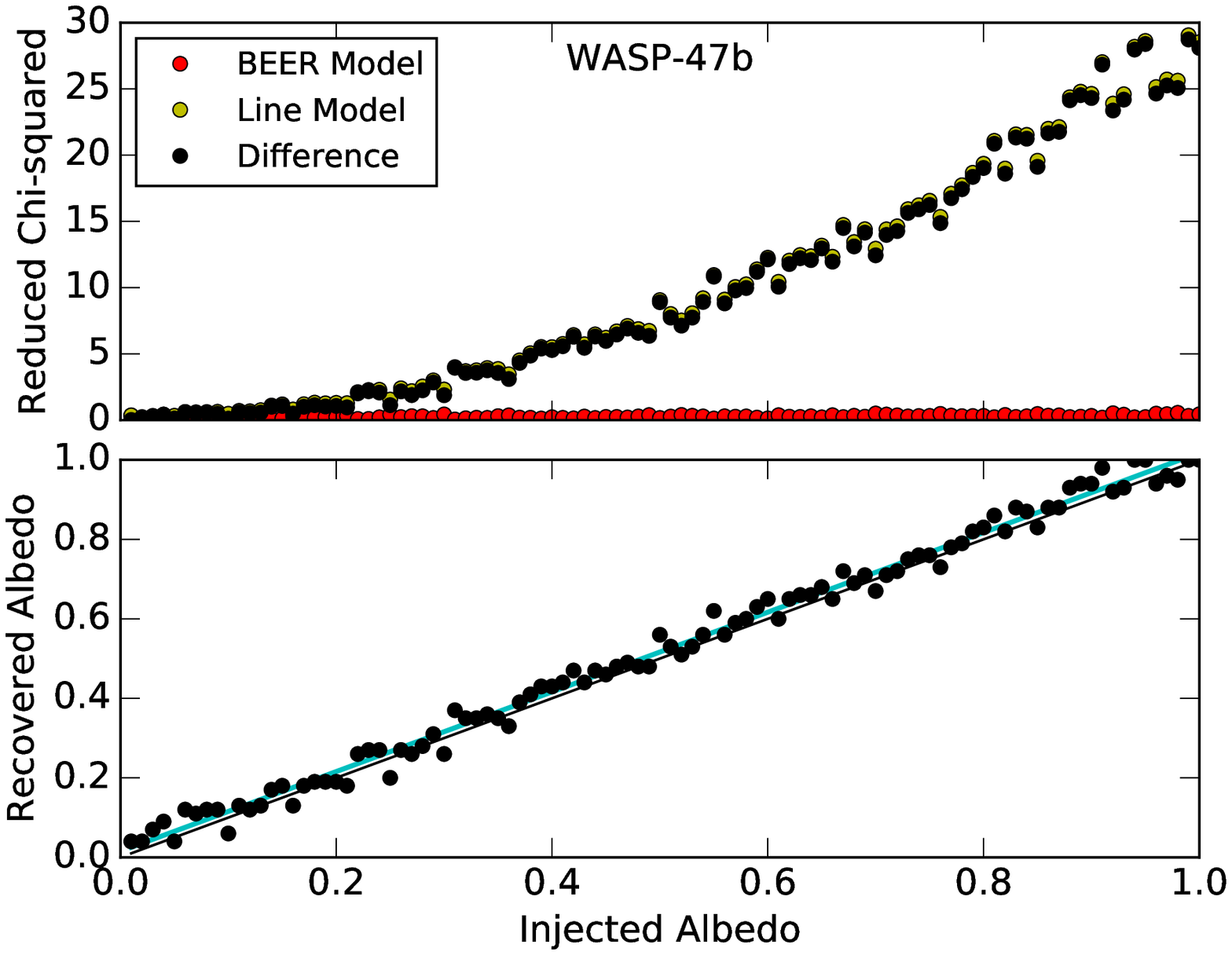} \\
    \includegraphics[width=8.5cm]{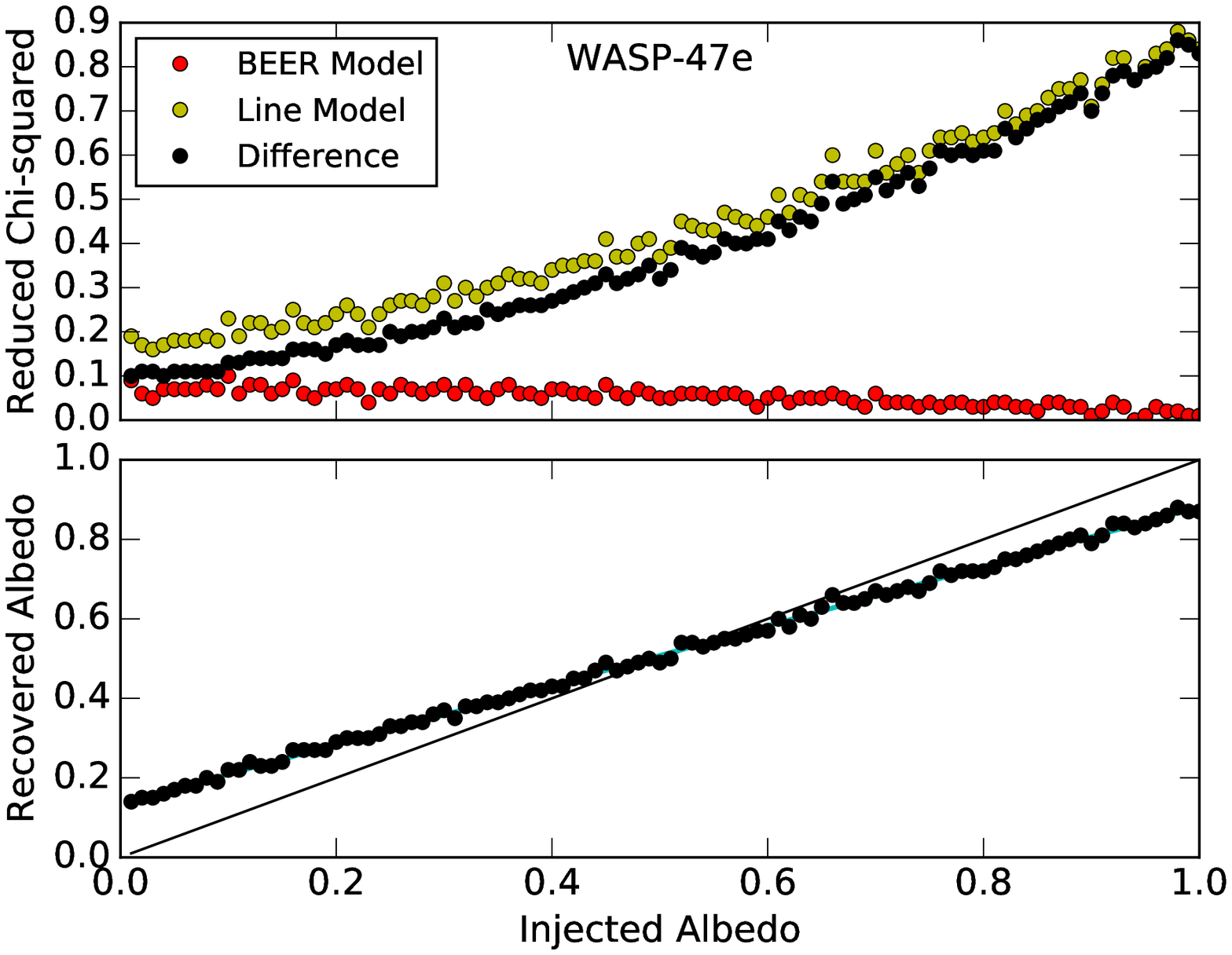}
  \end{center}
  \caption{The results of the phase variation analysis of the WASP-47
    b (top two panels) and e (bottom two panels) planets. For each of
    the two planets, the top panel indicates the reduced chi-squared
    for the best-fit BEER model (red points), a horizontal line model
    (yellow points), and their difference (black points). The values
    for the reduced chi-squared of the two models are offset
    vertically for clarity. The bottom panel shows a comparison
    between the injected albedo and the recovered (measured)
    albedo. The thin solid black line shows where the recovered
    albedos equal the injected albedos. For the b planet, the bottom
    panel also includes a thick light blue line underneath the black
    points that is fit to the recovered versus injected albedos. The
    thick light blue line emphasizes the vertical offset from equally
    recovered and injected albedos, which we interpret as a tentative
    0.016 geometric albedo for WASP-47b.}
  \label{chi2fig}
\end{figure}


\subsection{Results}
\label{phaseresults}

We find that the phase curve injections in steps of 0.01 albedo result
in increasingly higher recovered albedo measurements, as expected, for
the b and e planets. The WASP-47d phase curve, on the other hand,
remains inconsistent with the BEER model based on the significance and
signs of the measured semi-amplitudes, even when the maximum albedo
($A_g=1$) is injected.  In order to quantify the significance of the
models that were fit to the injected phase curve photometry, we
measure the difference between the reduced chi-squared ($\chi_\nu^2$)
of a horizontal line at the average flux of the data
($\chi^2_{\nu\text{,line}}$) and the reduced chi-squared of the
best-fit BEER model ($\chi^2_{\nu\text{,BEER}}$). Figure~\ref{chi2fig}
shows the reduced chi-squared of the two models, their difference, and
the extracted albedo as a function of the injected albedo for the b
and e planets of WASP-47.

\begin{figure*}
  \begin{center}
    \begin{tabular}{cc}
      \includegraphics[clip,width=8.5cm]{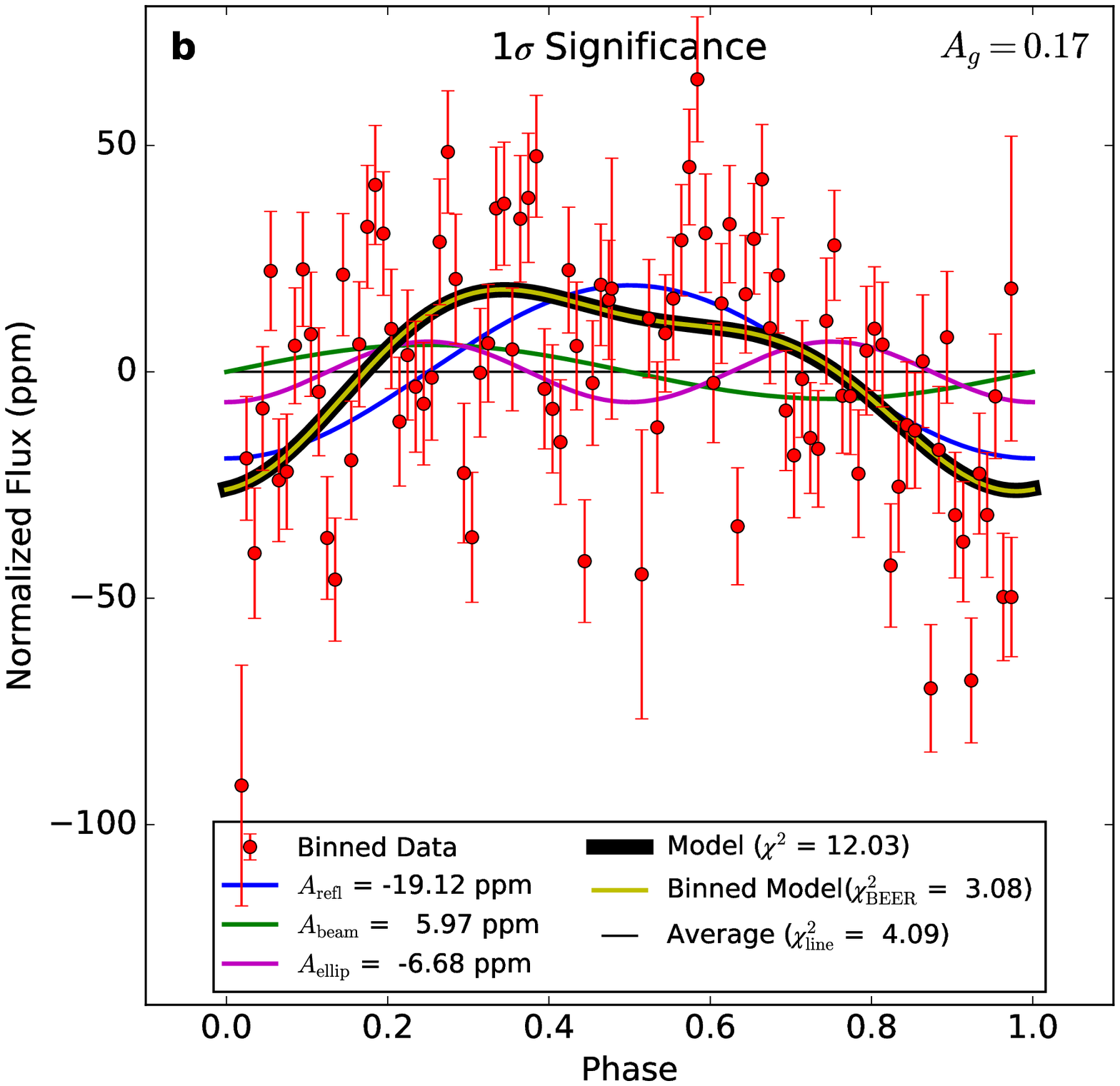} &
      \includegraphics[clip,width=8.5cm]{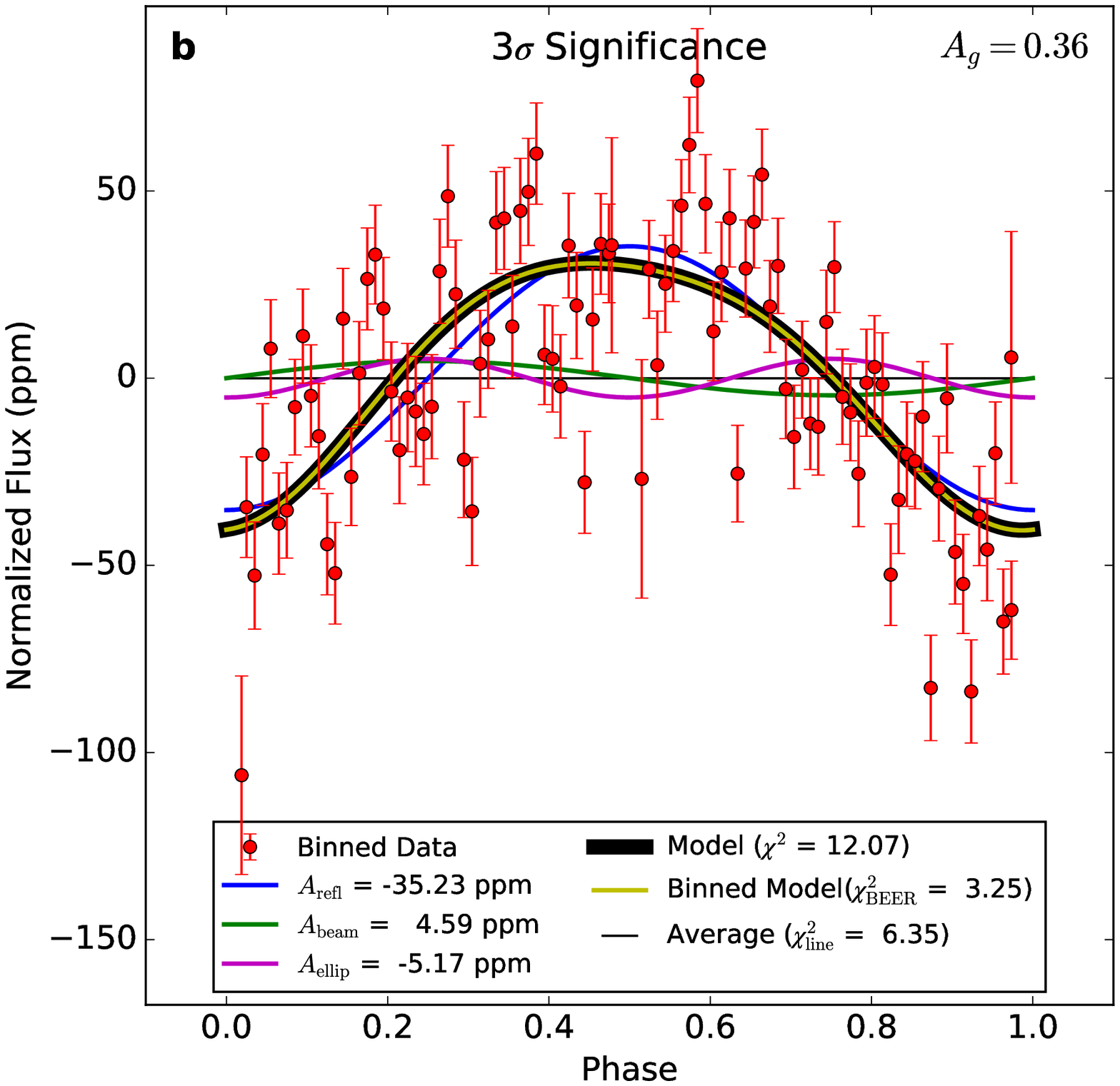} \\
      \includegraphics[clip,width=8.5cm]{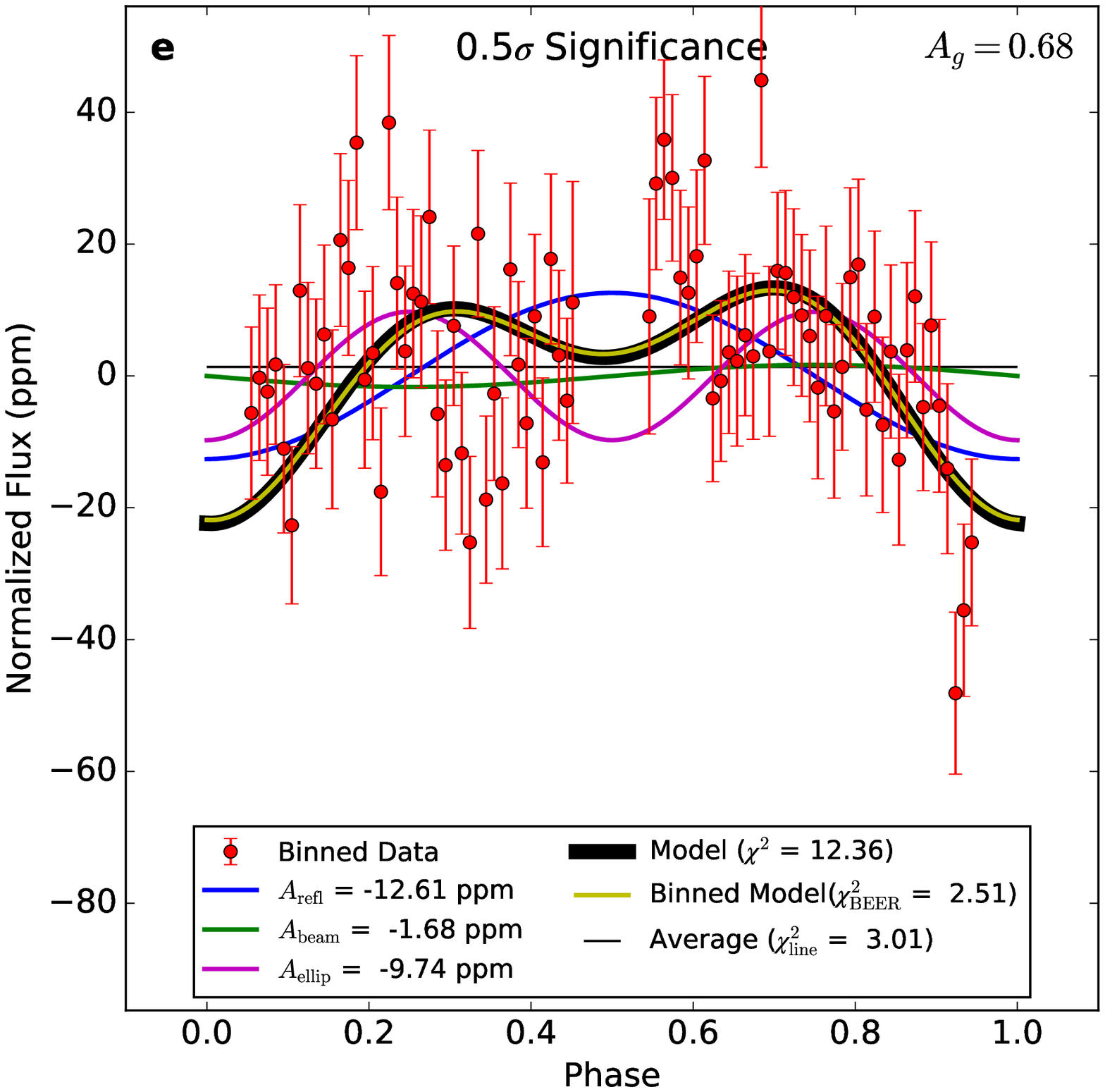} &
      \includegraphics[clip,width=8.5cm]{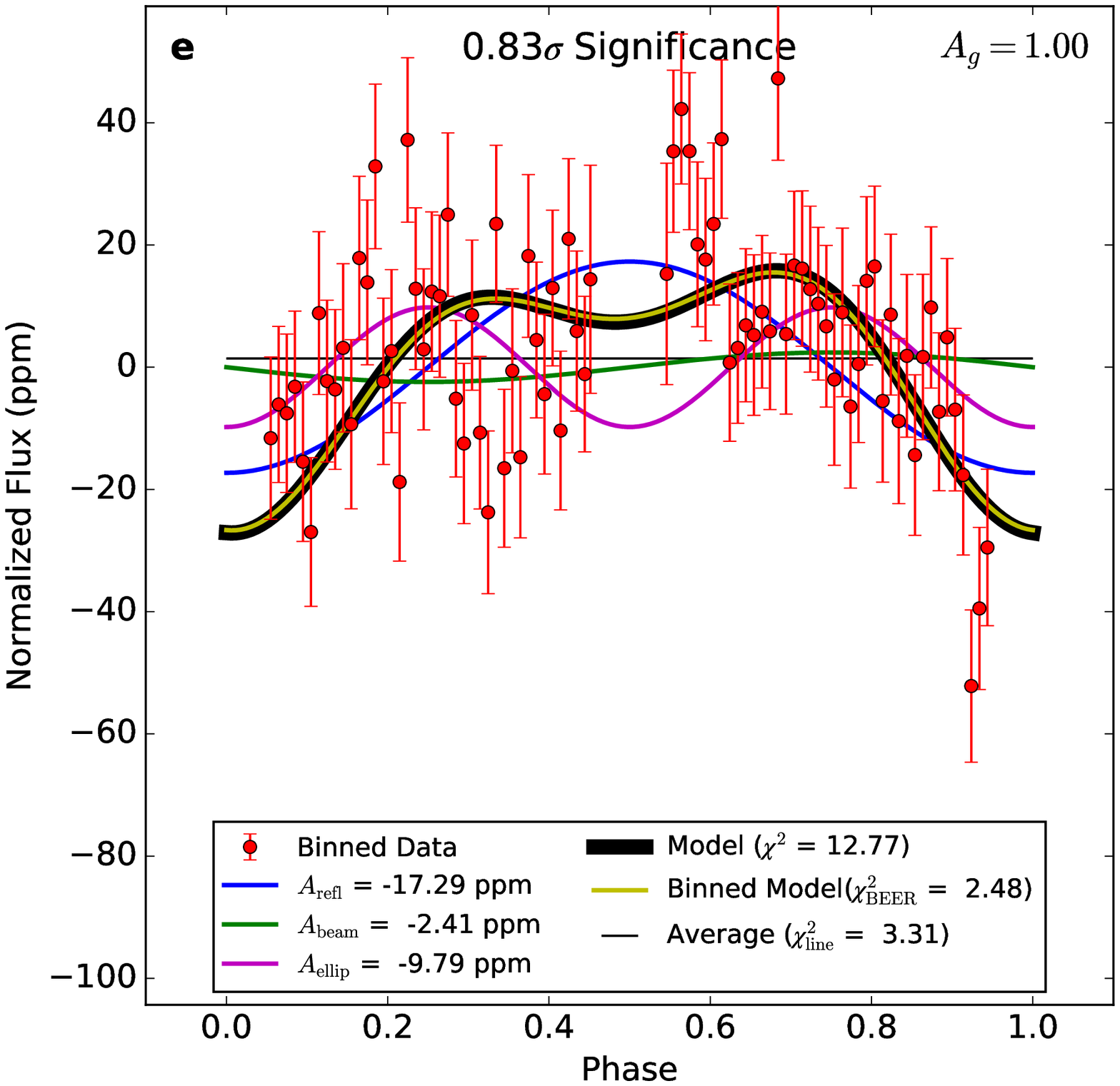} \\
    \end{tabular}
  \end{center}
  \caption{Results of the BEER model phase curve injection tests (see
    Sections~\ref{injtest} and \ref{phaseresults}) that were used to
    infer upper limits on the albedos of the b (top row) and e (bottom
    row) planets in WASP-47. The red points show the binned {\it K2}
    photometry that has been injected with the expected phase curve
    for a given geometric albedo, which is listed in the top right
    corner of each panel. The thick black and thin yellow curves show
    the best-fit BEER models to the unbinned (not shown) and binned
    (red points) photometry, respectively. The individual components
    of the BEER model are shown by the blue (reflection), green
    (beaming), and purple (ellipsoidal) curves. The thin horizontal
    black line indicates the average flux of the data and is used to
    compare the phase curve to a flat line.  The significance of the
    best-fit model compared to a flat line is listed at the top of
    each panel. All corresponding injected and recovered
    semi-amplitudes, albedos, and reduced chi-squared values are
    reported in Table~\ref{injtab}.}
  \label{beerfig}
\end{figure*}

As can be seen by the top panels of Figure~\ref{chi2fig}, the
extracted albedos for WASP-47b are incredibly consistent with the
injected albedos. Furthermore, there is an underlying offset in the
extracted albedos, such that the extracted albedos are on average
0.016 higher than the injected albedos. This offset could be caused by
each injected phase curve being added on top of the underlying phase
curve of WASP-47b. Since all BEER effects are left as free parameters
in the injection test phase curve fitting, the injected BEER effects
are assumed to be added to the intrinsic phase curve that was
undetected in Figure~\ref{original}. Therefore, based on the offset
between the injected and recovered albedos in Figure~\ref{chi2fig}, we
infer a potential geometric albedo of 0.016 for WASP-47b.  However,
this tentative 0.016 albedo is less than the 0.026 scatter of the
extracted albedos. We infer $1\sigma$ and $3\sigma$ upper limits on
the geometric albedo of WASP-47b based on the significance of the
measured BEER phase curve from the injection tests. We define
statistical significance using the difference between the reduced
chi-squared for the horizontal line and BEER models
($\chi^2_{\nu\text{,diff}}=\chi^2_{\nu\text{,line}}-\chi^2_{\nu\text{,BEER}}$).
We use a $\chi^2_{\nu\text{,diff}}$ of 1 and 3 to infer that the
$1\sigma$ and $3\sigma$ upper limits of the WASP-47b geometric albedo
are 0.17 and 0.36, respectively. The semi-amplitudes, albedos, and
reduced chi-squared values of the injected and recovered phase curves
based on the 0.17 and 0.36 albedo injection tests for WASP-47b are
given in Table~\ref{injtab} and the respective best-fit phase curves
are shown in the top panels of Figure~\ref{beerfig}.

Initially, the recovered albedo measurements for WASP-47e were
systematically lower than the injected albedos. The intrinsic phase
curve of WASP-47e cannot explain the {\it lower} albedo measurements.
The phase variations of WASP-47e may be undetectable or removed when
correcting the {\it K2} photometry, especially considering that the
realignment of the {\it K2} spacecraft occurred approximately on the
same timescale as the orbit of the e planet. The phase curve fit is
also sensitive to the input transit duration time.  In addition to
removing the expected transits and occultations, the transit duration
time also affects how much of the out-of-transit light curve is
included in the fit near 0 and 0.5 phase. These phases also represent
minima and maxima of the reflection and ellipsoidal variations, such
that the inclusion of in-transit photometry or the exclusion of too
much of the out-of-transit photometry could affect the measured
semi-amplitudes of these effects. We found that reducing the transit
duration time by 15~minutes ($t_{14} = 1.7$~hr) results in the best
recovered albedos compared to the injected albedos. The reduced
chi-squared statistics and the recovered versus injected albedos are
shown in the bottom panels of Figure~\ref{chi2fig}. Oddly, there is a
0.75 slope between the extracted and injected albedos for WASP-47e.
We suspect that the discrepancy between the extracted and injected
albedos could be attributed to (1) the injected phase curve being
altered by the corrections to the {\it K2} photometry that are on
approximately the same timescale as the orbital period of the e
planet, (2) the injected phase curve being added to the underlying
intrinsic phase curve, or (3) a combination thereof. Regardless, we
cannot constrain the geometric albedo of WASP-47e as we did for
WASP-47b, since the difference between the reduced chi-squared of the
horizontal line and BEER models never exceeds unity (i.e.,
$\chi^2_{\nu\text{,diff}}$ is always $<$1). Therefore, the injected
phase curve photometry and the best-fit recovered phase curves shown
in the bottom panels of Figure~\ref{beerfig} are based on the
injection tests that result from (1) the best-fit BEER phase curve
being detected at a 0.5$\sigma$ significance that infers an albedo of
0.68, and (2) injecting a BEER phase curve with a maximum geometric
albedo of $A_g = 1.0$ (0.83$\sigma$ significance). The
semi-amplitudes, albedos, and reduced chi-squared values of the
respective injected and recovered phase curves are given in
Table~\ref{injtab}.


\subsection{Potential Blended Phase Curves}
\label{multiphase}

For multi-planet systems with planets that are in resonant orbits,
there may be concern that a strong phase variation signature of one
planet may be blended with the phase curves of other planets in the
system. None of the known planets in WASP-47 are in resonance with
each other, but for completeness we investigate whether a strong
injected phase curve of an individual planet influences the phase
curves of other planets in the system. Since the phase curve of
WASP-47e is not significantly detected ($<$1$\sigma$) at an albedo of
1, we inject a phase curve with an unphysically high geometric albedo
equal to 10 for all of the planets together, then the b, d, and e
planets individually. For each of these injection cases, we repeat our
phase curve analysis on the b, d, and e planets to see how a strong
reflection modulation may influence the phase curves of the other
planets in the system.

The shape of the WASP-47d phase curve remains generally the same in
all test cases, but is most significantly modified by a strong
injected phase curve of the b planet. This is because the b planet has
the strongest phase curve signature of all the planets in the WASP-47
system and it is near a 2:1 resonance with the d planet. However, in
all cases the WASP-47d phase curve remains inconsistent with a BEER
model based on the significance and sign of the measured semi-amplitudes. 
Since WASP-47b exhibits the strongest phase variation
signature, the changes to its phase curve are negligible regardless of
whether the contributions from the phase curves of the other planets
are included. Curiously, when only the WASP-47e phase curve is
injected, the recovered albedo is {\it lower} than when all phase
curves are included. Once again, the extracted albedo is lower than
the injected albedo for WASP-47e, suggesting that the intrinsic phase
curve is suppressed by the low signal-to-noise or the corrections of
the {\it K2} photometry that are approximately on the same timescale 
as the orbit of the e planet.


\section{Discussion}
\label{discussion}

As described in Section~\ref{phase}, analysis of potential exoplanet
phase signatures have been carried out by numerous groups. In many
cases, the albedo measurements, or upper limits, have led to the
conclusion that hot Jupiters tend to have very low albedos
\citep{rowe2008,bell2017,mocnik2018,mallonn2019b} due to significant
constraints on the presence of icy condensates in the atmospheres of
such planets. Note that \citet{rowe2008} and \citet{mocnik2018}
establish the low albedo estimates through analysis of the
out-of-transit/eclipse phase variations rather than just eclipse
measurements, similar to the methodology we present here. There exist
exceptions to low albedos for hot Jupiters however, such as Kepler-7b
\citep{demory2011a}, no doubt reflecting the diversity of possible
atmospheric compositions. Our results for WASP-47b are consistent with
the majority of cases where the planets are demonstrated to have
poorly reflective atmospheres. The Neptune-size WASP-47d planet lies
well below the detection capabilities of the {\it K2} data, as
predicted by the calculations shown in Table~\ref{paramtab}, but may
also have a blended phase curve due to a near 2:1 resonance with the
stronger phase curve of the b planet. In the case of the e planet, the
calculations of Table~\ref{paramtab} indicate that the phase signature
could be detected if the planet is highly reflective, such as a molten
surface. Our analyses in Section~\ref{k2} show that the e planet phase
curve lies at the threshold of {\it K2} photometric detectability,
possibly due to photometric corrections for the periodic
($\sim$1\,day) pointing alignment of the spacecraft being on a similar
timescale as the orbit of the e planet ($P = 0.79$\,days). The
diversity of geometric albedos is expected to be higher for
terrestrial planets due to the diversity of surfaces and
atmospheres. Hot terrestrial planets like WASP-47e may have a limited
variety of reflective properties since the high incident flux can
potentially erode atmospheres
\citep{tian2009b,roettenbacher2017,zahnle2017,brogi2018b} as well as
create molten surface environments.

We investigated the possibility of additional photometry of WASP-47
from other facilities, such as the Transiting Exoplanet Survey
Satellite ({\it TESS}), described in detail by
\citet{ricker2015}. WASP-47 has ecliptic coordinates of 328.996$\degr$
longitude and -0.209$\degr$ latitude. Being so close to the ecliptic,
the star does not coincide with {\it TESS} observations during the
primary mission, though it will likely be observed in extended mission
scenarios that target the ecliptic region of the sky
\citep{sullivan2015}. Furthermore, WASP-47 is bright enough to serve
as an excellent target for follow-up observations using the
CHaracterising ExOPlanets Satellite ({\it CHEOPS}), which was launched
in early 2020 \citep{broeg2014}. Observations of WASP-47 using both
{\it TESS} and {\it CHEOPS} brings a significant advantage in
multi-wavelength observations of atmospheric phase signatures since,
as noted in Section~\ref{phase}, the reflected light component is
wavelength dependant \citep{gaidos2017b}. Note also that the most
significant correction to the {\it K2} photometry described in
Section~\ref{k2} is for the unstable pointing, which is on the order
of $\sim$1\,day and is similar to the orbital period of WASP-47e. This
further emphasizes the advantage of follow-up observations from a more
stable spacecraft such as {\it TESS} or {\it CHEOPS}.

The dynamical evolution of the known planets in the WASP-47 system is,
on its own, a fascinating aspect of the system. The compact nature of
the known inner three planets is relatively unperturbed by the
presence of planet c in an eccentric orbit within the HZ. As described
in Section~\ref{system}, further acquisition of RV data may reveal an
additional planet in a longer period orbit, well outside of the HZ. If
so, the farther giant planet could be regularly exchanging angular
momentum through oscillation of eccentricities, as has been observed
in other systems \citep{kane2014b}. Given the discovery history of the
system, it is likely that there are yet further planet discoveries to
be made in the system with additional RV monitoring.

WASP-47c is a gas giant and so is not deemed habitable on its own, but
it does have the potential to be the host of large rocky exomoons that
would also be in the HZ. An exomoon would have the benefit of a
diversity of sources providing energy to their potential biosphere,
not just a reliance on the flux received from the host star. The
reflected light and emitted heat of the host planet as well as tidal
heating forces caused by the motion of the moon orbiting the planet
will provide additional energy to a moon
\citep{heller2013a,hinkel2013b}. These combined heating effects
effectively extend the HZ for these moons, creating a wider temperate
area in which they may maintain conditions on their surface that are
amenable to liquid surface water \citep{scharf2006}. For a planet like
WASP-47c that spends part of its orbit near the outer edge of the HZ
(see Figure~\ref{systemfig}), these extra energy sources may enable
potential exomoons in orbit to maintain habitable conditions during
apastron.

While there have been preliminary detections of exomoon signatures
\citep{teachey2018a,teachey2018b} no exomoons have been confirmed to
date \citep{heller2019b,kreidberg2019a}. However, there is a general
consensus in the community that the existence of exomoons is likely
\citep{williams1997a,kipping2009c,heller2012,heller2015d,zollinger2017}
and it has even been proposed that there could be as many or even more
exomoons in the HZ as there are terrestrial planets \citep{hill2018}.

Any potentially habitable moon would need to have a large enough mass
to maintain an atmosphere that can support liquid surface water, and
may need a magnetosphere to protect its atmosphere and surface from
the radiation emitted by the host planet and potentially needs
tectonic plates to facilitate carbon cycling \citep{williams1997a}. It
has been posed by \citet{heller2013a} that a moon would need to be at
least $\geq 0.25M\oplus$ to maintain these conditions.  Using the
relationship derived in \citet{canup2002}, the maximum mass of a moon
that is formed in~situ with WASP-47~c is
$\sim$0.04~$M_\oplus$. Therefore it is likely that for a moon to be
considered habitable about this planet, it would need to have been
captured rather than formed in~situ with the planet
\citep{williams2013}.


\section{Conclusions}
\label{conclusions}

The WASP-47 system has been described as ``the gift that keeps on
giving,'' with numerous exoplanet discoveries beyond the initial
discovery of the hot Jupiter, WASP-47b. The orbital architecture of
the system that combines compact orbits with a giant planet in an
eccentric orbit in the HZ presents an interesting dynamical
configuration. As discussed in this paper, the HZ giant planet is a
potential abode for exomoons and may represent a common scenario of
giant planet migration into the HZ that can truncate HZ terrestrial
planet occurrence rates \citep{hill2018}. The eccentric nature of the
giant planet orbit may be indicative of past planet-planet scattering
\citep{carrera2019b} or an ongoing interaction with a more distant,
yet undiscovered, planet \citep{kane2014b}. The complete orbital
architecture remains open with the likelihood of at least one
additional planet in the system.

However, the most prominent source of new discoveries within the
system may originate from atmospheric studies of the inner transiting
planets of the system. The architecture of an inner super-Earth
followed by a gas giant planet has been observed in other systems,
such as 55~Cancri \citep{kane2011f}, and may be related to planet
formation processes in compact systems \citep{weiss2018a}. This
combination of a relatively small inner planet with a gas giant
neighbor is compelling from a phase variation point of view since the
planets can have similar phase amplitudes. Our calculations have shown
that planet e with a lava surface can have a comparable phase
amplitude to planet b if that planet is of particularly low
albedo. Our analysis of the {\it K2} photometry demonstrates that
planet b must have a very low albedo, consistent with albedo
measurements and/or constraints for other short period giant
planets. The phase signature of planet e lies barely beneath the
statistically significant detection threshold of the {\it K2}
photometry, but our injection test and subsequent analysis shows that
the difference in reduced chi-squared for the BEER model diverges for
a broad range of injected albedos.

The plethora of dark planets in exoplanetary systems, such as
WASP-47b, will present an interesting sample from which to study the
relationship between atmospheric composition and geometric
albedo. Systems such as WASP-47 and those similar systems with bright
host stars discovered with {\it TESS} will thus provide a concise
target set from which to form the basis of a transmission spectroscopy
study with the {\it James Webb Space Telescope} \citep{kempton2018}
and may also be possible for terrestrial targets such as WASP-47e
\citep{batalha2018b}. It is thus expected that the continued
monitoring of WASP-47 and uncovering architecturally similar systems
has a great deal more to teach us about the planet formation,
evolution, and planetary atmospheres.


\section*{Acknowledgements}

The authors would like to thank Juliette Becker and Andrew Vanderburg
for guidance regarding the {\it K2} photometry. Thanks are also due to
the anonymous referee, whose comments greatly improved the quality of
the paper. This research has made use of the following archives: the
Habitable Zone Gallery at hzgallery.org and the NASA Exoplanet
Archive, which is operated by the California Institute of Technology,
under contract with the National Aeronautics and Space Administration
under the Exoplanet Exploration Program. The results reported herein
benefited from collaborations and/or information exchange within
NASA's Nexus for Exoplanet System Science (NExSS) research
coordination network sponsored by NASA's Science Mission Directorate.


\software{RadVel \citep{fulton2018a}}




\end{document}